\documentclass[aps,prl,reprint,longbibliography]{revtex4-1}
\usepackage{appendix}
\usepackage{bm}
\usepackage{enumerate}
\usepackage{graphicx}
\usepackage{dcolumn}
\usepackage{bm}

\usepackage{amssymb}
\usepackage{amsmath}
\usepackage{booktabs}
\usepackage{braket}
\usepackage[caption = false]{subfig}

\usepackage{tikz}
\usetikzlibrary{patterns}
\tikzset{
    cross/.pic = {
    \draw[rotate = 45] (-#1,0) -- (#1,0);
    \draw[rotate = 45] (0,-#1) -- (0, #1);
    }
}
\usetikzlibrary {patterns.meta}
\pgfdeclarepattern{
      name=hatch,
      parameters={\hatchsize,\hatchangle,\hatchlinewidth},
      bottom left={\pgfpoint{-.1pt}{-.1pt}},
      top right={\pgfpoint{\hatchsize+.1pt}{\hatchsize+.1pt}},
      tile size={\pgfpoint{\hatchsize}{\hatchsize}},
      tile transformation={\pgftransformrotate{\hatchangle}},
      code={
        \pgfsetlinewidth{\hatchlinewidth}
        \pgfpathmoveto{\pgfpoint{-.1pt}{-.1pt}}
        \pgfpathlineto{\pgfpoint{\hatchsize+.1pt}{\hatchsize+.1pt}}
        \pgfpathmoveto{\pgfpoint{-.1pt}{\hatchsize+.1pt}}
        \pgfpathlineto{\pgfpoint{\hatchsize+.1pt}{-.1pt}}
        \pgfusepath{stroke}
      }
    }

\tikzset{
      hatch size/.store in=\hatchsize,
      hatch angle/.store in=\hatchangle,
      hatch line width/.store in=\hatchlinewidth,
      hatch size=3pt,
      hatch angle=0pt,
      hatch line width=.5pt,
    }

\newcommand{\ee}{{\rm e}}
\newcommand{\ii}{{\rm i}}

\tikzset{every picture/.style={line width=0.75pt}} 

\begin{document}

\preprint{APS/123-QED}

\title{Gapless Edge Gravitons and Quasiparticles in Fractional Quantum Hall Systems with Non-Local Confinement}

\author{Daniel Spasic-Mlacak  and Nigel R. Cooper}

\affiliation{
T.C.M Group, Cavendish Laboratory, University of Cambridge, J.J. Thompson Avenue, Cambridge CB3 0US, United Kingdom \looseness=-1}
\date{\today}

\begin{abstract}

One of the central tenets of the theory of the fractional quantum Hall effect is that the bulk quantized Hall response requires the existence of a gapless chiral edge mode.
The field theoretical arguments for this rely on locality. While locality is typically met in standard experimental settings, it need not always apply. 
Motivated by experimental capabilities of photonic platforms, we study confining potentials that are step-like in angular momentum, and thus non-local in position.
We show that this non-local potential does not host conventional chiral edge modes. These are replaced by gapless spin-2 edge states, which we show are connected to the collective ‘graviton’ excitations that are gapped in the bulk.
Furthermore, we show that FQH states host  gapless (charged) quasiparticles on their edges, even in the absence of conventional edge modes.
The edge state energies vanish as a power-law in system size, with an exponent that characterises the bulk topological order.

\end{abstract}

\maketitle

In the fractional quantum Hall (FQH) effect, the low-energy excitations at the boundary are described by the chiral Luttinger liquid, a one-dimensional system where interactions lead to non-Fermi liquid behavior and power-law tunneling characteristics. The chiral Luttinger liquid theory of edge states has played an important role in advancing the understanding of the bulk-edge correspondence in topologically ordered systems \cite{wen1995topological}. Experimentally, edge state measurements have served as powerful probes of the underlying bulk topological order \cite{grayson1998continuum, de1998direct}. 

Bulk excitations have also been central to the study of the incompressible FQH fluid. The seminal work of Girvin, MacDonald, and Platzman (GMP) introduced the magnetoroton mode, a gapped neutral collective excitation arising from density fluctuations \cite{girvin1985collective}. More recently, Haldane's geometric description of FQH states has highlighted the importance of an emergent internal metric, which characterizes local deformations of the quantum Hall fluid \cite{haldane2011geometrical}. This geometric degree of freedom is intimately connected to the structure of the magnetoroton mode. It has been theoretically linked to an emergent spin-2 collective excitation, often referred to as the `graviton', which has been observed experimentally via inelastic light scattering~\cite{liang2024evidence}. 

In this paper, we elucidate a connection between the bulk and edge excitations of FQH states,  by studying a `non-local hard-wall' confining potential which is step-like in the angular momentum basis. Such a confining potential has been proposed in the context of photonic platforms\cite{umucalilar2021autonomous}, where polaritons in twisted optical cavities have been used to realise the two-body Laughlin state \cite{clark2020observation}.  
This potential differs from the usually considered `hard-wall' potentials, for which the energy is a smoothly increasing function of single particle angular momentum for large particle numbers  \cite{cooper2015signatures,fern2017quantum,macaluso2017hard, macaluso2018ring}. Being discontinuous in the angular momentum basis, the potential we consider is necessarily {\it non-local} in the position basis.

\tikzset{every picture/.style={line width=0.75pt}} 
\begin{figure} [htp]
\centering
\begin{tikzpicture}[x=0.5pt,y=0.5pt,yscale=-0.8,xscale=0.8]

\draw [line width=1.5]    (82.67,189.67) -- (112.67,189.67);
\draw [line width=1.5]    (132.67,189.67) -- (162.67,189.67) ;
\draw [line width=1.5]    (182.67,189.67) -- (212.67,189.67) ;
\draw [line width=1.5]    (232.67,189.67) -- (262.67,189.67) ;
\draw [line width=1.5]    (282.67,189.67) -- (312.67,189.67) ;
\draw [line width=1.5]    (332.67,89.67) -- (362.67,89.67) ;
\draw [line width=1.5]    (382.67,89.67) -- (412.67,89.67) ;
\draw [line width=1.5]    (432.67,89.67) -- (462.67,89.67) ;
\draw [line width=1.5]    (482.67,89.67) -- (512.67,89.67) ;
\draw    (550.82,90) -- (549.84,188) ;
\draw [shift={(549.82,190)}, rotate = 270.57] [color={rgb, 255:red, 0; green, 0; blue, 0 }  ][line width=0.75]    (10.93,-3.29) .. controls (6.95,-1.4) and (3.31,-0.3) .. (0,0) .. controls (3.31,0.3) and (6.95,1.4) .. (10.93,3.29)   ;
\draw    (549.82,190) -- (550.8,92) ;
\draw [shift={(550.82,90)}, rotate = 90.57] [color={rgb, 255:red, 0; green, 0; blue, 0 }  ][line width=0.75]    (10.93,-3.29) .. controls (6.95,-1.4) and (3.31,-0.3) .. (0,0) .. controls (3.31,0.3) and (6.95,1.4) .. (10.93,3.29)   ;
\draw  [fill={rgb, 255:red, 0; green, 0; blue, 0 }  ,fill opacity=1 ] (143,189.67) .. controls (143,187) and (145.09,184.83) .. (147.67,184.83) .. controls (150.24,184.83) and (152.33,187) .. (152.33,189.67) .. controls (152.33,192.34) and (150.24,194.5) .. (147.67,194.5) .. controls (145.09,194.5) and (143,192.34) .. (143,189.67) -- cycle ;
\draw  [fill={rgb, 255:red, 0; green, 0; blue, 0 }  ,fill opacity=1 ] (193,189.67) .. controls (193,187) and (195.09,184.83) .. (197.67,184.83) .. controls (200.24,184.83) and (202.33,187) .. (202.33,189.67) .. controls (202.33,192.34) and (200.24,194.5) .. (197.67,194.5) .. controls (195.09,194.5) and (193,192.34) .. (193,189.67) -- cycle ;
\draw  [fill={rgb, 255:red, 0; green, 0; blue, 0 }  ,fill opacity=1 ] (243,189.67) .. controls (243,187) and (245.09,184.83) .. (247.67,184.83) .. controls (250.24,184.83) and (252.33,187) .. (252.33,189.67) .. controls (252.33,192.34) and (250.24,194.5) .. (247.67,194.5) .. controls (245.09,194.5) and (243,192.34) .. (243,189.67) -- cycle ;
\draw  [fill={rgb, 255:red, 0; green, 0; blue, 0 }  ,fill opacity=1 ] (293,189.67) .. controls (293,187) and (295.09,184.83) .. (297.67,184.83) .. controls (300.24,184.83) and (302.33,187) .. (302.33,189.67) .. controls (302.33,192.34) and (300.24,194.5) .. (297.67,194.5) .. controls (295.09,194.5) and (293,192.34) .. (293,189.67) -- cycle ;
\draw  [fill={rgb, 255:red, 0; green, 0; blue, 0 }  ,fill opacity=1 ] (93,189.67) .. controls (93,187) and (95.09,184.83) .. (97.67,184.83) .. controls (100.24,184.83) and (102.33,187) .. (102.33,189.67) .. controls (102.33,192.34) and (100.24,194.5) .. (97.67,194.5) .. controls (95.09,194.5) and (93,192.34) .. (93,189.67) -- cycle ;

\draw    (297.67,259.01) -- (297.67,230.93) ;
\draw [shift={(147.67+3*50,228.93)}, rotate = 90.25] [color={rgb, 255:red, 0; green, 0; blue, 0 }  ][line width=0.75]    (10.93,-3.29) .. controls (6.95,-1.4) and (3.31,-0.3) .. (0,0) .. controls (3.31,0.3) and (6.95,1.4) .. (10.93,3.29)   ;
\draw    (197.67,259) -- (297.67,259.01) ;
\draw    (197.67,259) -- (197.67,200) ;
\draw [color={rgb, 255:red, 208; green, 2; blue, 27 }  ,draw opacity=1 ][line width=1.5]    (257.67,268.67) -- (237.67,248.67) ;
\draw    (347.67,51.83) -- (347.67,78.83) ;
\draw [shift={(347.67,80.83)}, rotate = 270] [color={rgb, 255:red, 0; green, 0; blue, 0 }  ][line width=0.75]    (10.93,-3.29) .. controls (6.95,-1.4) and (3.31,-0.3) .. (0,0) .. controls (3.31,0.3) and (6.95,1.4) .. (10.93,3.29)   ;
\draw    (247.67,52) -- (347.67,51.83) ;
\draw    (247.67,52) -- (247.67,175.83) ;
\draw [color={rgb, 255:red, 208; green, 2; blue, 27 }  ,draw opacity=1 ][line width=1.5]    (257.67,248.67) -- (237.67,268.67) ;
\draw    (297.67,32) -- (397.67,31.83) ;
\draw    (297.67,32) -- (297.67,175.83) ;
\draw    (397.67,31.83) -- (397.67,78.83) ;
\draw [shift={(397.67,80.83)}, rotate = 269.74] [color={rgb, 255:red, 0; green, 0; blue, 0 }  ][line width=0.75]    (10.93,-3.29) .. controls (6.95,-1.4) and (3.31,-0.3) .. (0,0) .. controls (3.31,0.3) and (6.95,1.4) .. (10.93,3.29)   ;

\draw (287,207.4) node [anchor=north west][inner sep=0.75pt]    {$M$};
\draw (562,131.4) node [anchor=north west][inner sep=0.75pt]    {$V_{\text{conf}}$};

\draw [line width=1,-stealth]    (69,200) -- (600,200);
\draw [line width=0.5]    (97.67,200) -- (97.67,205);
\draw [line width=0.5]    (147.67,200) -- (147.67,205);
\draw [line width=0.5]    (147.67+50,200) -- (147.67+50,205);
\draw [line width=0.5]    (147.67+2*50,200) -- (147.67+2*50,205);
\draw [line width=0.5]    (147.67+3*50,200) -- (147.67+3*50,205);
\draw [line width=0.5]    (147.67+4*50,200) -- (147.67+4*50,205);
\draw [line width=0.5]    (147.67+5*50,200) -- (147.67+5*50,205);
\draw [line width=0.5]    (147.67+6*50,200) -- (147.67+6*50,205);
\draw [line width=0.5]    (147.67+7*50,200) -- (147.67+7*50,205);
\draw (590,220) node  {$m$};

\end{tikzpicture}
\caption{Schematic diagram showing the free-fermion ground state in the non-local potential, with all states $m\leq M$ occupied, and states $m>M$ unoccupied with energy $V_{\text{conf}}$. Any excitation involves moving a particle to an orbital with $m>M$, and thus the spectrum is gapped. (Here illustrated for $\Delta L =2$.)}
\label{tikzdensity}
\end{figure}
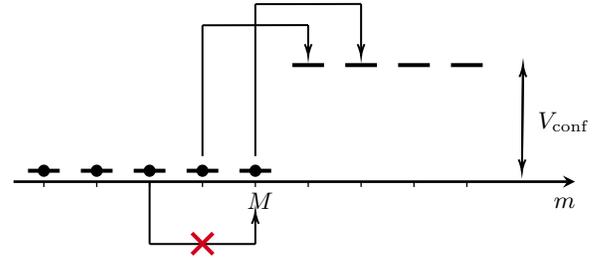

It is natural to wonder how the phenomenology of edge states is modified by this non-local potential.
The general result that a bulk system described by a Chern-Simons field theory must have a chiral edge mode on a manifold with a boundary\cite{witten1989quantum,wen1991} cannot be applied, because that field-theoretical approach relies on locality.
Is there a gapless edge state at all? 
Certainly the $\nu=1$ state of non-interacting fermions loses its gapless edge state: in the ground state the fermions fill all of the states with zero potential, so any excitation must move an electron to an orbital of a nonzero potential $V_{\rm conf}$. (See Fig.~\ref{tikzdensity}.) 
But what about fractional quantum Hall states? As we will show, even for $V_{\rm conf}\to \infty $ the FQH states in this non-local confinement do exhibit gapless states in the thermodynamic limit $N\to \infty$. However, these gapless states are not conventional edge modes. Rather they are connected to the bulk graviton modes, and associated `parent' quasiparticles.
Furthermore, we show that the parent quasiparticles themselves are gapless for $N\to\infty$, establishing that the {\it edges} of FQH states are compressible even in the absence of conventional edge modes.


We consider the Laughlin states at filling factors $\nu=1/q$ \cite{laughlin1983anomalous}. We work exclusively in the lowest Landau level (LLL), where the single-particle wavefunctions $\phi_{m}(\textbf{r})$ are indexed by angular momentum $m=0,1,2,\ldots$. The non-local confining potential is then 
\begin{equation}
\label{MainConfinement}
\hat{V}_{M} = V_{\text{conf}} \sum_{m>M} \hat{c}^{\dagger}_{ m} \hat{c}_{m}
\end{equation}
where $M = q (N-1)$ is the highest occupied orbital of the Laughlin state and the operators $\hat{c}_{m}$ annihilate bosons (fermions) with angular momentum $m$ and obey the appropriate (anti)commutation relations. 
(For the real-space form of the non-local potential see the Supplementary Material.) 
The inter-particle interaction may be written as $\hat{U}_{\text{int}} = \sum_{m} \sum_{i<j} {\cal V}_{m} {\cal P}_{m}(ij)$, where ${\cal V}_{m}$ are the Haldane pseudopotentials, ${\cal P}_{m}(ij)$ projects to the many-body state where particles $i$ and $j$ have relative angular momentum $m$ (which is even/odd for bosons/fermions). To make the Laughlin state at $\nu=1/q$ the exact zero-energy groundstate, we choose ${\cal V}_m = {\cal V}_{\rm int} \geq 0 $ for $m\leq q-2$ and ${\cal V}_m=0 $ otherwise.  

\begin{figure}[htp]
\centering
\includegraphics[width=3.4in]{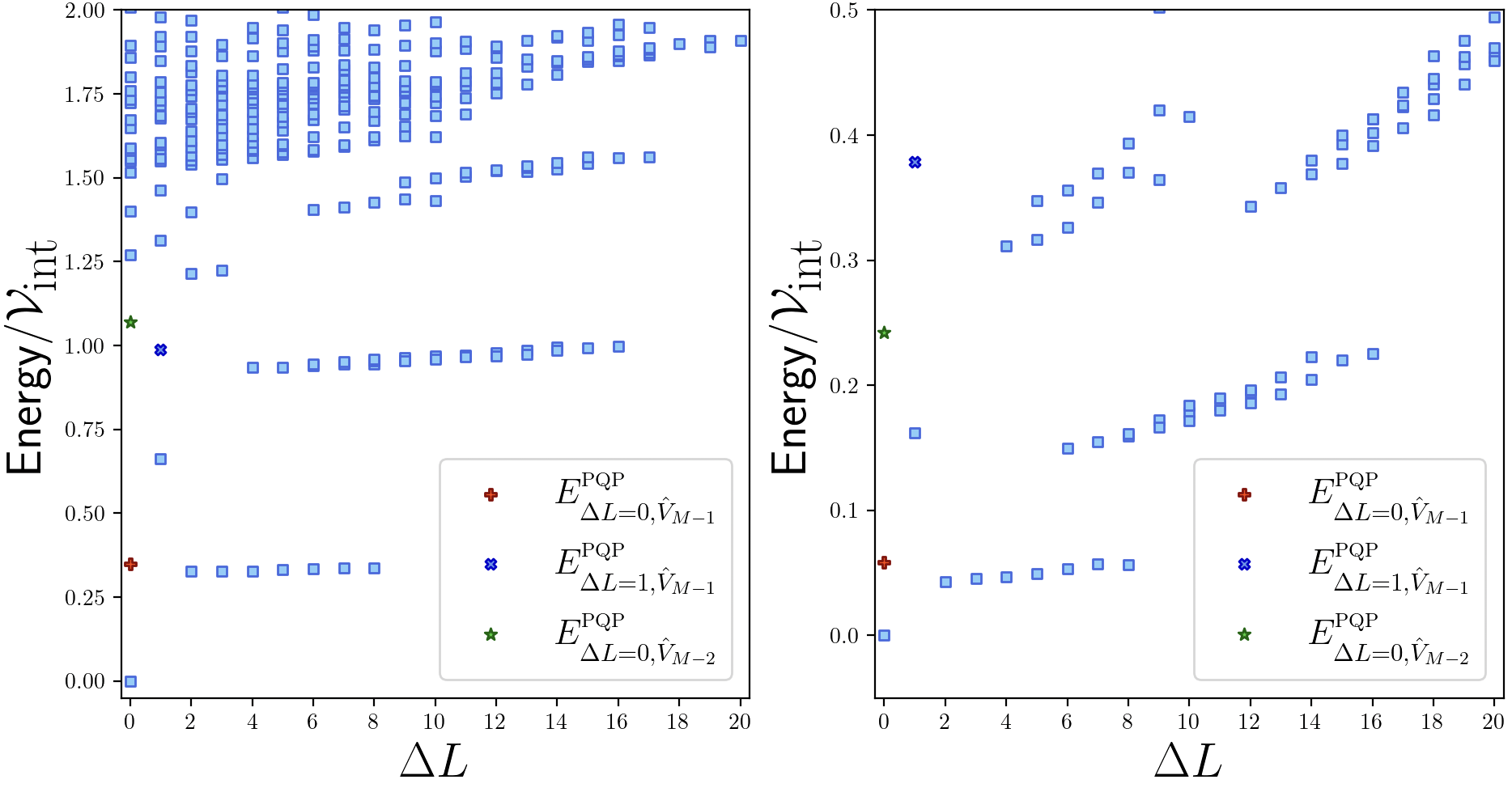}
\caption{Energy spectra in the strong confinement limit for $N=8$, $\nu=1/2$ bosons (left) and $\nu=1/3$ fermions (right). The spectrum of edge states is made of branches of bound states, sharply contrasting the chiral Luttinger liquid theory. Lowest energies for confining potentials $\hat{V}_{M-1}$, $\hat{V}_{M-2}$ at $\Delta L=0,1$ are used to illustrate the connection of branch states to their associated `parent' quasiparticles (PQPs). }
\label{SCLBosons+Fermions}
\end{figure}

We numerically investigated the excitation spectrum above the zero energy Laughlin ground state. Exact diagonalisation was performed  
for $N$ particles in the potential $\hat{V}_{M}$ and interactions $\hat{U}_{\text{int}}$, using the DiagHam library \cite{DiagHam}. 
We focus first on the strong confinement limit (SCL), where states with $m>M$ are excluded from the basis and the only energy scale is the interaction energy ${\cal V}_{\rm int}$. We will later also mention results in the weak confinement limit (WCL) $V_{\text{conf}} \ll {\cal V}_{\rm int}$, insofar as it shares interesting connections with the SCL. 

The energy spectra for edge excitations of $N=8$, $\nu={1}/{2}$ bosons and $\nu=1/3$ fermions in the SCL are shown in Figure \ref{SCLBosons+Fermions}, where the energy is plotted as a function of $\Delta L = L -\frac{1}{2} NM $, 
chosen such that the  Laughlin ground state is at $\Delta L=0$.
%

\begin{figure}[htp]
\centering
\includegraphics[width=3.4in]{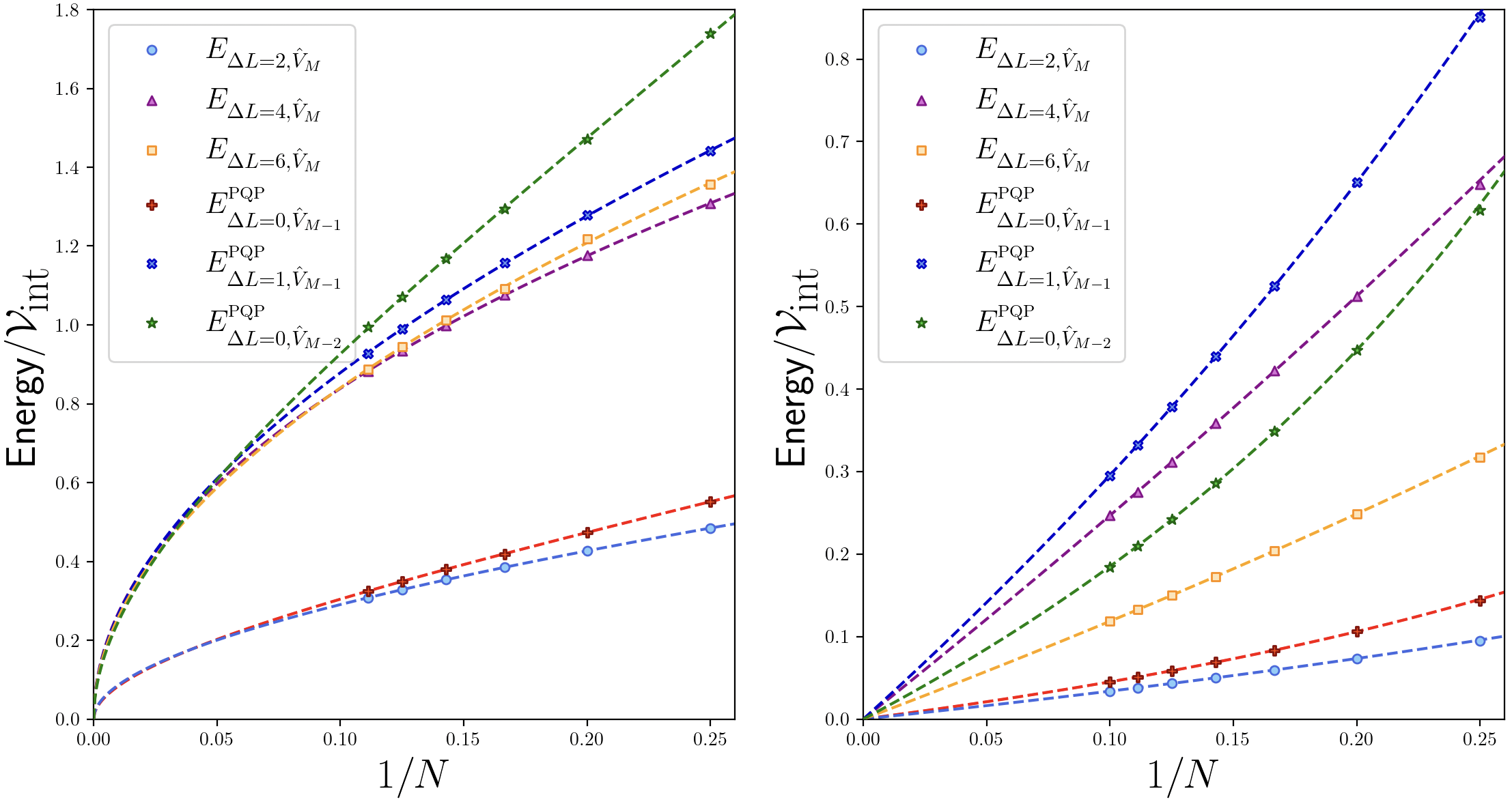}
\caption{Energies of branch modes and `parent' quasiparticles as a function of system size for $\nu=1/2$ bosons (left) and $\nu=1/3$ fermions (right). The extrapolation to zero in the thermodynamic limit is done by fitting to $E = E_{0}\frac{N^{-1/2}}{1-AN^{-1/2}}$ (bosons) and $E =E_{0}\frac{N^{-1}}{1-AN^{-1}}$ (fermions), according to the algebraic exponent $\frac{1-q}{2}$ in Eq.~(\protect\ref{SCLVariational}).}
\label{ThermoBosons+Fermions}
\end{figure}

The excitation spectrum shows several striking features.
(i) The low energy branch is not of conventional ``edge mode" character, in which the energy is a linear function of $\Delta L$ with a slope determined by the edge mode velocity.
The energy is not even monotonically increasing with $\Delta L$, with $\Delta L = 1$ having an anomalously large energy. 
(ii) Instead an apparent band of low energy states appears from $\Delta L =2, \ldots, N$. with very weak dispersion.
(iii) This entire band of states has an energy that vanishes as a power-law in $1/N$ in the thermodynamic limit $N\to \infty$ (see Fig.~\ref{ThermoBosons+Fermions}).
In the following we will explain the nature of this unconventional low-energy spectrum, and will derive an analytic expression that motivates the power-law fits shown in Fig.~\ref{ThermoBosons+Fermions}.

To describe the nature of the many-body states, it is useful to work in a subspace of squeezed partition states generated from a specified root partition \cite{bernevig2008model}. The root partition indicates the dominant occupation number configuration of angular momentum orbitals, with the convention that angular momentum $m$ increases from left to right. For example, the root partition $\{1011\}$ corresponds to particles occupying angular momenta $m=0,2,3$. Other physically relevant states are generated via a squeezing operation, which conserves total angular momentum $L$ and reflects the generalised Pauli exclusion principle of FQH states. A generic state in the squeezed subspace can be constructed as a sum over all monomials ${\cal M}_{\mu}$ for which the partitions $\mu$ are squeezed from $\lambda$, denoted $\mu \preceq \lambda$.
 For example, the $\nu=\frac{1}{2}$ Laughlin state has root configuration  $\lambda=\{101010\ldots101\}$ and the coefficients of squeezed configurations $\mu\preceq \lambda$ can be found recursively [they correspond to the Jack polynomial $J^{-2}_{\lambda}(\{\textbf{r}_{i}\})$]. 

\tikzset{every picture/.style={line width=0.75pt}} 
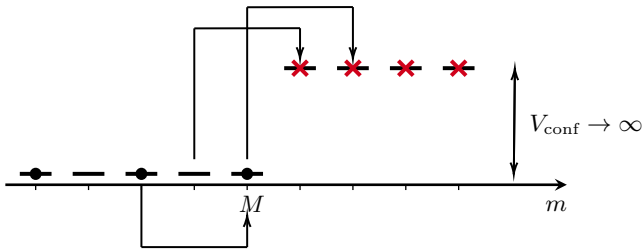
\begin{figure} [htp]
\centering
\begin{tikzpicture}[x=0.5pt,y=0.5pt,yscale=-0.8,xscale=0.8]

\draw [line width=1.5]    (82.67,189.67) -- (112.67,189.67);
\draw [line width=1.5]    (132.67,189.67) -- (162.67,189.67) ;
\draw [line width=1.5]    (182.67,189.67) -- (212.67,189.67) ;
\draw [line width=1.5]    (232.67,189.67) -- (262.67,189.67) ;
\draw [line width=1.5]    (282.67,189.67) -- (312.67,189.67) ;
\draw [line width=1.5]    (332.67,89.67) -- (362.67,89.67) ;
\draw [line width=1.5]    (382.67,89.67) -- (412.67,89.67) ;
\draw [line width=1.5]    (432.67,89.67) -- (462.67,89.67) ;
\draw [line width=1.5]    (482.67,89.67) -- (512.67,89.67) ;
\draw    (550.82,90) -- (549.84,188) ;
\draw [shift={(549.82,190)}, rotate = 270.57] [color={rgb, 255:red, 0; green, 0; blue, 0 }  ][line width=0.75]    (10.93,-3.29) .. controls (6.95,-1.4) and (3.31,-0.3) .. (0,0) .. controls (3.31,0.3) and (6.95,1.4) .. (10.93,3.29)   ;
\draw    (549.82,190) -- (550.8,92) ;
\draw [shift={(550.82,90)}, rotate = 90.57] [color={rgb, 255:red, 0; green, 0; blue, 0 }  ][line width=0.75]    (10.93,-3.29) .. controls (6.95,-1.4) and (3.31,-0.3) .. (0,0) .. controls (3.31,0.3) and (6.95,1.4) .. (10.93,3.29)   ;

\draw  [fill={rgb, 255:red, 0; green, 0; blue, 0 }  ,fill opacity=1 ] (93,189.67) .. controls (93,187) and (95.09,184.83) .. (97.67,184.83) .. controls (100.24,184.83) and (102.33,187) .. (102.33,189.67) .. controls (102.33,192.34) and (100.24,194.5) .. (97.67,194.5) .. controls (95.09,194.5) and (93,192.34) .. (93,189.67) -- cycle ;

\draw  [fill={rgb, 255:red, 0; green, 0; blue, 0 }  ,fill opacity=1 ] (193,189.67) .. controls (193,187) and (195.09,184.83) .. (197.67,184.83) .. controls (200.24,184.83) and (202.33,187) .. (202.33,189.67) .. controls (202.33,192.34) and (200.24,194.5) .. (197.67,194.5) .. controls (195.09,194.5) and (193,192.34) .. (193,189.67) -- cycle ;

\draw  [fill={rgb, 255:red, 0; green, 0; blue, 0 }  ,fill opacity=1 ] (293,189.67) .. controls (293,187) and (295.09,184.83) .. (297.67,184.83) .. controls (300.24,184.83) and (302.33,187) .. (302.33,189.67) .. controls (302.33,192.34) and (300.24,194.5) .. (297.67,194.5) .. controls (295.09,194.5) and (293,192.34) .. (293,189.67) -- cycle ;
\draw    (297.67,259.01) -- (297.67,230.93) ;
\draw [shift={(147.67+3*50,228.93)}, rotate = 90.25] [color={rgb, 255:red, 0; green, 0; blue, 0 }  ][line width=0.75]    (10.93,-3.29) .. controls (6.95,-1.4) and (3.31,-0.3) .. (0,0) .. controls (3.31,0.3) and (6.95,1.4) .. (10.93,3.29)   ;
\draw    (197.67,259) -- (297.67,259.01) ;
\draw    (197.67,259) -- (197.67,200) ;

\draw    (347.67,51.83) -- (347.67,78.83) ;
\draw [shift={(347.67,80.83)}, rotate = 270] [color={rgb, 255:red, 0; green, 0; blue, 0 }  ][line width=0.75]    (10.93,-3.29) .. controls (6.95,-1.4) and (3.31,-0.3) .. (0,0) .. controls (3.31,0.3) and (6.95,1.4) .. (10.93,3.29)   ;
\draw    (247.67,52) -- (347.67,51.83) ;
\draw    (247.67,52) -- (247.67,175.83) ;

\draw    (297.67,32) -- (397.67,31.83) ;
\draw    (297.67,32) -- (297.67,175.83) ;
\draw    (397.67,31.83) -- (397.67,78.83) ;
\draw [shift={(397.67,80.83)}, rotate = 269.74] [color={rgb, 255:red, 0; green, 0; blue, 0 }  ][line width=0.75]    (10.93,-3.29) .. controls (6.95,-1.4) and (3.31,-0.3) .. (0,0) .. controls (3.31,0.3) and (6.95,1.4) .. (10.93,3.29)   ;

\draw [color={rgb, 255:red, 208; green, 2; blue, 27 }  ,draw opacity=1 ][line width=1.5]    (354.67,96.67) -- (340.67,82.67) ;
\draw [color={rgb, 255:red, 208; green, 2; blue, 27 }  ,draw opacity=1 ][line width=1.5]    (354.67,82.67) -- (340.67,96.67) ;

\draw [color={rgb, 255:red, 208; green, 2; blue, 27 }  ,draw opacity=1 ][line width=1.5]    (404.67,96.67) -- (390.67,82.67) ;
\draw [color={rgb, 255:red, 208; green, 2; blue, 27 }  ,draw opacity=1 ][line width=1.5]    (404.67,82.67) -- (390.67,96.67) ;

\draw [color={rgb, 255:red, 208; green, 2; blue, 27 }  ,draw opacity=1 ][line width=1.5]    (454.67,96.67) -- (440.67,82.67) ;
\draw [color={rgb, 255:red, 208; green, 2; blue, 27 }  ,draw opacity=1 ][line width=1.5]    (454.67,82.67) -- (440.67,96.67) ;

\draw [color={rgb, 255:red, 208; green, 2; blue, 27 }  ,draw opacity=1 ][line width=1.5]    (504.67,96.67) -- (490.67,82.67) ;
\draw [color={rgb, 255:red, 208; green, 2; blue, 27 }  ,draw opacity=1 ][line width=1.5]    (504.67,82.67) -- (490.67,96.67) ;

\draw (287,207.4) node [anchor=north west][inner sep=0.75pt]    {$M$};
\draw (562,131.4) node [anchor=north west][inner sep=0.75pt]    {$V_{\text{conf}} \rightarrow \infty$};

\draw [line width=1,-stealth]    (69,200) -- (600,200);
\draw [line width=0.5]    (97.67,200) -- (97.67,205);
\draw [line width=0.5]    (147.67,200) -- (147.67,205);
\draw [line width=0.5]    (147.67+50,200) -- (147.67+50,205);
\draw [line width=0.5]    (147.67+2*50,200) -- (147.67+2*50,205);
\draw [line width=0.5]    (147.67+3*50,200) -- (147.67+3*50,205);
\draw [line width=0.5]    (147.67+4*50,200) -- (147.67+4*50,205);
\draw [line width=0.5]    (147.67+5*50,200) -- (147.67+5*50,205);
\draw [line width=0.5]    (147.67+6*50,200) -- (147.67+6*50,205);
\draw [line width=0.5]    (147.67+7*50,200) -- (147.67+7*50,205);
\draw (590,220) node  {$m$};

\end{tikzpicture}
\caption{Schematic diagram of the root configuration of lowest energy $\Delta L=2$ excitation of the confining potential (\ref{MainConfinement}) in the SCL for $\nu=1/2$ bosons, starting from the root configuration of the Laughlin state, $\{\ldots 1,0,1,0,1\}$. As angular momentum orbitals above $M$ are forbidden by the confinement, the $\Delta L=2$ excitation closest to the edge involves moving an additional particle to the $M$ orbital to give $\{\ldots 1,0,0,0,2\}$.}
\label{SchematicSCR}
\end{figure}

The low-energy branch has states that range from $\Delta L =2, \ldots N$, which we can use to find the appropriate root partitions. In the SCL, we are constrained to working in $M+1$ possible angular momentum orbitals ($m=0,...,M$). To reproduce the right counting of the lowest branch, we postulate root partitions with two quasiparticles and two quasiholes on the edge for $\Delta L=2$, and other states in the branch corresponding to moving one of the two quasiholes in the bulk (which there is precisely $N-2$ ways to do). For example, the $\nu=1/2$ Laughlin state of $N=4$ particles has root partition $\{1010101\}$, and its $\Delta  L = 2$ edge state root partition is given by $\{1010002\}$.  The root partitions for the other states in the lowest branch corresponding to $\Delta L = 3, 4$ are then $\{1001002\}$ and $\{0101002\}$ respectively. 

We have verified that the states in Figure \ref{SCLBosons+Fermions} are constructed from the proposed root partitions by calculating the projection of these states onto the subspace of squeezed states $\bra{\psi}\hat{P}_{\mu \preceq \lambda} \ket{\psi}$. Doing so, we find that states in the restricted subspace are a very good approximation for the whole $\Delta L = 2$ branch. (For example this overlap is $\gtrsim 0.99992$ for $N=9$.)  

We note that these root partitions are identical to the configurations of the bulk magnetoroton mode on the sphere~\cite{yang2012model}. The $\Delta L=2$ state then corresponds to the spin-2 graviton of the FQHE, which is now on the edge. (See Fig.~\ref{SchematicSCR}.)
Thus, the non-local potential acts to trap the branch of graviton modes on the edge, lowering its energy well below the bulk gap. As we now show, the residual non-zero energy of this trapped graviton mode is, in fact, a finite-size effect that vanishes as $N\to \infty$.


The very weak dispersion of the low-energy branch, with $\Delta L = 2,\ldots N$ motivates the view that one can think of this as a (mobile) quasihole that interacts very weakly with a `quasielectron' that is bound to the edge. Within this picture, the branch can be related to a fundamental `parent' quasiparticle excitation. For example, for $\nu=1/2$ bosons this is the quasiparticle with root partition $\{101\ldots1002\}$. This root partition state has been studied in the context of bulk excitations as the `quasielectron' of the Laughlin state and can be constructed using composite fermion theory \cite{yang2014nature, jeon2003nature}. On the edge, the state can be easily generated as the $\Delta L=0$ groundstate of a modified confinement in which the $m=M$ orbital is also excluded, i.e. $\hat{V}_{M-1}$. 
We find that the energy of this parent quasiparticle state is very close to that of its corresponding branch (see Figure \ref{SCLBosons+Fermions}). Under the assumption that the parent quasiparticle can lower its energy by binding with a free quasi-hole, then the parent quasiparticle energy provides an upper bound on the energy of the branch.

We now turn to show analytically that the energy of the parent quasiparticle vanishes in the thermodynamic limit.
As a variational state we take $\ket{\Psi^{\text{SCL}}_{\Delta L=0}} \propto (1-\hat{n}_{M})\ket{\Psi_{\rm L}}$, which removes the components of the Laughlin state with a particle in the $M$ orbital to make a zero energy eigenstate of $\hat{V}_{M-1}$. (Here $\hat{n}_{m}=\hat{c}^{\dagger}_{m}\hat{c}_{m}$.) The energy of this state is then given by
\begin{equation}
\label{EnergyTrialSCL}
E^{\text{SCL}}_{\Delta L=0} = \left(\frac{\bra{\Psi_{\rm L}} \hat{n}_{M}\hat{U}_{\text{int}}\hat{n}_{M}\ket{\Psi_{\rm L}}}{{\bra{\Psi_{\rm L}} \hat{n}_{M}\hat{n}_{M}\ket{\Psi_{\rm L}}}}\right)\left(\frac{\braket{\hat{n}_{M}}}{1-\braket{\hat{n}_{M}}}\right),
\end{equation}
where we have used the properties of the Laughlin state $\hat{U}_{\text{int}} \ket{\Psi_{\rm L}} =0$ and $\hat{n}_{M}^{2}\ket{\Psi_{\rm L}} = \hat{n}_{M}\ket{\Psi_{\rm L}}$, and $\braket{\hat{n}_{M}}\equiv\braket{\Psi_{\rm L}|\hat{n}_{M}|\Psi_{\rm L}}$. 
The first factor on the right hand side of (\ref{EnergyTrialSCL}) can be viewed as the interaction energy of one particle frozen in the single particle state $\phi_M(\textbf{r})$ with the mean density $\rho_{{\rm L},N-1}(\textbf{r})$ of the remaining Laughlin state of $N-1$ particles, i.e. $\int \rho_{{\rm L},N-1} (\textbf{r}) V_{\rm int}(\textbf{r}-\textbf{r}')|\phi_M(\textbf{r}')|^2 d^2 rd^2 r'$. Using the fact that this gives an interaction energy density of order ${\cal V}_{\rm int}$ (up to a numerical factor that can depend on $\nu$) within the region of the (normalized) state $\phi_M(\textbf{r})$,  the first factor must be of order ${\cal V}_{\rm int}$ for large $N$. (See Supplementary material for numerical evaluation.) The occupation number $\langle \hat{n}_M\rangle$  in the second factor can be directly related to the normalisation constant of the $\nu=1/q$ Laughlin state \cite{mitra1993angular}, for which an asymptotic expansion as a function of particle number is known \cite{di1994laughlin}, $\braket{\hat{n}_{M}} \sim f(q) N^{\frac{1-q}{2}}$ where $f(q)$ is some unknown function with $f(1)=1$. Putting these together,  we find to leading order
\begin{equation}
\label{SCLVariational}
E^{\text{SCL}}_{\Delta L=0}\sim  {\cal V}_{\rm int}f(q)N^{\frac{1-q}{2}}\,.
\end{equation} 
Thus, the gap vanishes with an exponent that depends on the nature of the bulk topological phase. This algebraic exponent $(1-q)/2$ in Eq.~(\ref{SCLVariational}) accurately fits the numerical results for the excitation spectrum,  Fig.~\ref{ThermoBosons+Fermions}. 

This variational result establishes that, for the non-local potential, the Laughlin ground state is gapless to the formation of a charged quasiparticle at $N\to \infty$. %
The FQH state has long been argued to be incompressible through numerical and experimental observations \cite{girvin1985collective, duncan1987hierarchy, pinczuk1993observation, morf2002excitation}, although a proof of a gap in the thermodynamic limit remains unknown \cite{rougerie2019laughlin}. These novel edge states give an unexpected case where quasiparticle excitations are not gapped, as opposed to their bulk counterparts. 
It also demonstrates that the {\it edge} of the Laughlin state is compressible even in the absence of conventional edge modes, which involve (small) fluctuations of particles into the confining potential.

The approach of associating a `parent' quasiparticle to the branch of edge excitations can be generalised to higher branches. For example, we find that the second branch for $N=8$ bosons has a (near) degeneracy pattern which can be explained by two overlapping branches: one spanning from $\Delta L=4$ to $\Delta L=9$, and the other from $\Delta L=6$ to $\Delta L=16$. The latter has a pattern of near degenerate states $1,1,2,2,3,3,3,2,2,1,1$ which can be reproduced by taking a root partition $\{101\ldots1000003\}$ and moving two quasiholes around the bulk (i.e. keeping $\{0003\}$ at the end of the partition). To obtain an analogous quasiparticle, we take the lowest energy state of $\hat{V}_{M-2}$ at $\Delta L=0$, which gives the parent quasiparticle $\{101\ldots1003\}$. From the numerics, the branch starting at $\Delta L=4$ must have one state for each $\Delta L$, so it is natural to conjecture the root configuration $\{101\ldots1000102\}$ and allow only one quasihole to move in the bulk (i.e., keeping $\{00102\}$ at the end of the partition). The analogous quasiparticle can then be obtained by taking the lowest energy state of confinement $\hat{V}_{M-1}$ at $\Delta L=1$, which gives the parent quasiparticle $\{101\ldots100102\}$. 

To test whether the proposed parent quasiparticles are related to these branches, we compare their energies numerically for bosons and fermions. For fermions, the two quasiparticles have energies very close to the two branches which they were assumed to be related to according to the root partition arguments above. For bosons, the energies of $\braket{\hat{V}_{M-1}}_{\Delta L= 1}$ and $\braket{\hat{V}_{M-2}}_{\Delta L= 0}$ quasiparticles happen to be close to each other, causing the two branches to overlap (see Fig.~\ref{SCLBosons+Fermions}). The energies of these parent quasiparticles also vanish like those of the associated neutral branches for $N\to\infty$ (see Fig.~\ref{ThermoBosons+Fermions}).

\begin{figure}[htp]
\centering
\includegraphics[width=3.4in]{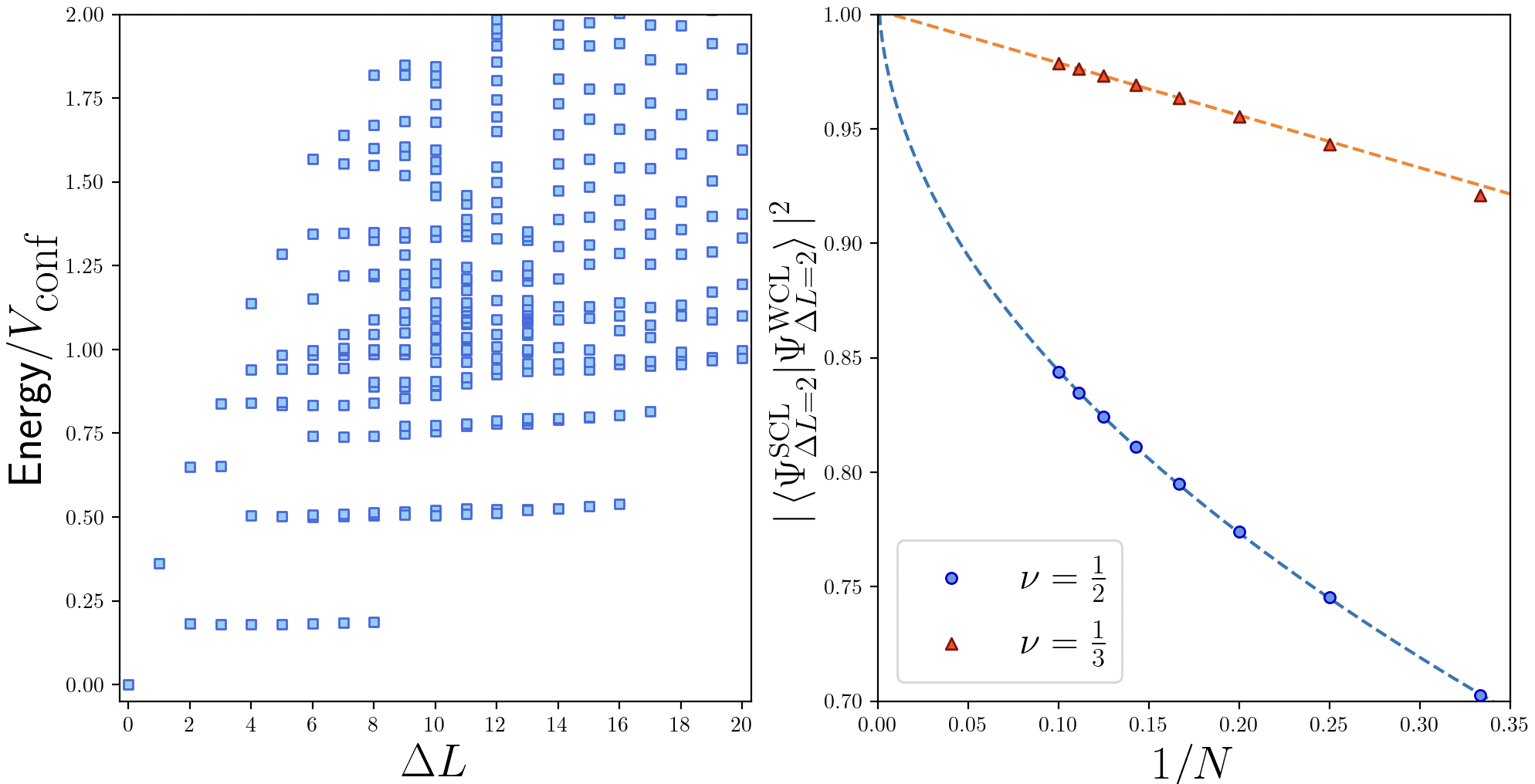}
\caption{(Left) Energy spectrum of edge states in the WCL for $N=8$, $\nu=1/2$ bosons. The branch structure corresponds exactly to that of the SCL. (Right) Thermodynamic behaviour of overlap squared of the $\Delta L =2$ state in the two limits. The two states are shown to have a very strong overlap which increases with system size. Applying a best fit with the critical exponent $\frac{1-q}{2}$ predicts remarkably well that the overlap approaches unity as $N\rightarrow \infty$ for both the bosonic and fermionic Laughlin state.}
\label{OverlapFigure}
\end{figure}

Finally, we turn to discuss the effect of the non-local potential in the WCL. The spectrum has many features in common with those of the SCL described above. 
Firstly, the energy spectrum in the WCL exhibits the same low-energy structure as the SCL, as shown in Fig.~\ref{OverlapFigure} (left). The same root configurations also describe the low-energy states. Indeed the wavefunctions in the WCL and SCL converge in the thermodynamic limit: see Fig.~\ref{OverlapFigure} (right), which shows that the wavefunction overlap approaches unity as $N\to \infty$ according to
$|\braket{\Psi^{\text{SCL}}_{\Delta L =2}|\Psi^{\text{WCL}}_{\Delta L =2}}|^{2} = 1-O(N^{\frac{1-q}{2}})$.   
Secondly, the energies of the low energy states go to zero in the thermodynamic limit, also with the same exponent as in the SCL. A proof of this is provided for general $\Delta L$ in the Supplemental Material. Here we will illustrate this for $\Delta L=1$, taking the variational state $\ket{\Psi^{\text{WCL}}_{\Delta L =1}} \propto  \hat{L}^\dag \ket{\Psi_{\rm L}}$,
where  $\ket{\Psi_{\rm L}}$ is the Laughlin state and 
we have defined the raising operator of angular momentum,
$\hat{L}^{\dagger} = \sum_{m=0}^{M} \sqrt{2(m+1)}\hat{c}^{\dagger}_{m+1}\hat{c}_{m}$.
The energy of the edge excitation can be written as
\begin{equation}
\label{EnergyTrialWCL}
E_{\Delta L = 1}^{\text{WCL}}=V_{\text{conf}}|\alpha_{M}|^{2}\frac{\braket{\hat{n}_{M}}}{\braket{\Psi_{\rm L}|\hat{L}\hat{L}^{\dagger}|\Psi_{\rm L}}},
\end{equation}
where the key simplification arises from the observation that $\hat{V}_{M}\hat{L}^{\dagger}\ket{\Psi_{\rm L}}= [\hat{V}_{M},\hat{L}^{\dagger}]\ket{\Psi_{\rm L}}$, as $\hat{V}_{M}\ket{\Psi_{\rm L}}=0$. 
The thermodynamic limit of the angular momentum occupation number $\braket{\hat{n}_{M}}$ was already discussed in the context of the SCL in (\ref{SCLVariational}). 
Finally, the denominator in (\ref{EnergyTrialWCL}) has a known closed form in the thermodynamic limit \cite{dubail2012edge} (see Supplementary Material for further details), giving 
\begin{equation}
\label{thermolimitenergy}
E_{\Delta L = 1}^{\text{WCL}} \sim V_{\text{conf}} [{f}(q)/q] N^{\frac{1-q}{2}}\,.
\end{equation} 
Hence, the energy scales to zero with the same exponent as in SCL (\ref{SCLVariational}), or also remains gapped for an IQH state with filling fraction $\nu = 1$. The additional factor of $1/q$ compared to equation (\ref{SCLVariational}) is also evident in the numerical results shown in Fig.~\ref{SCLBosons+Fermions} (left) and Fig.~\ref{OverlapFigure} (left).




In conclusion, we have shown that the non-local confinement (\ref{MainConfinement}) does not host conventional edge modes. Rather, it supports gapless modes that are directly related to the spin-2 graviton modes in the bulk. These edge excitations are linked to parent quasiparticles which have vanishing energy on the boundary. Spectroscopic measurements of the energy gap in this non-local confinement\cite{umucalilar2021autonomous}, as a function of system size, could be used to extract the algebraic exponent associated with the non-trivial edge structure. The exact diagonalisation results indicate that the exponent can be extracted in experiments using small particle numbers.  

\textit{Acknowledgements -- } 
We are grateful to Jon Simon and Duncan Haldane for stimulating discussions and feedback. This work was supported by the EPSRC [grant number EP/V062654/1], a Simons Investigator Award [Grant No. 511029] and a Cambridge International Scholarship provided by the Cambridge Trust. 
For the purpose of open access, the authors have applied a creative commons attribution (CC BY) licence to any author accepted manuscript version arising.

\nocite{*}

\bibliography{library.bib}

\section{Supplementary Material}

\section{Non-locality of confining potential}
To convert the confining potential (\ref{MainConfinement}) to position basis, we write
\begin{equation}
\begin{aligned}
\hat{V}_{M} = \int d^{2}\textbf{r}\int d^{2}\textbf{r}^{\prime} K(\textbf{r},\textbf{r}^{\prime})\psi^{\dagger}(\textbf{r})\psi(\textbf{r}^{\prime}),
\\
K(\textbf{r},\textbf{r}^{\prime})=V_{\rm conf} \sum_{m>M} \phi^{*}_{m}(\textbf{r}) \phi_{m}(\textbf{r}^{\prime}),
\end{aligned}
\end{equation}
where the operators $\psi(\textbf{r})$ annihilate bosons(fermions) at position $\textbf{r}$, and $\phi_{m}(\textbf{r}^{\prime})$ are the single-particle angular momentum states.
Inserting their explicit form into the expression for $K(\textbf{r},\textbf{r}^{\prime})$, we find 
\begin{equation}
\begin{aligned}
K(\textbf{r},\textbf{r}^{\prime})=\frac{V_{\rm conf}}{2\pi l_{B}^{2}} \exp\left(-\frac{r^{2}+{r^{\prime}}^{2}}{4l_{B}^{2}} +\ee^{\ii(\phi^{\prime}-\phi)} \frac{rr^{\prime}}{2l_{B}^{2}}\right) \\
\times[1-\frac{\Gamma(M+1,\frac{rr^{\prime}}{2l_{B}^{2}}\ee^{\ii(\phi^{\prime}-\phi)})}{\Gamma(M+1)}],
\end{aligned}
\end{equation}
where $\textbf{r}=(r,\phi)$ in polar co-ordinates. This is clearly a non-local confinement, and remains non-local even in the thermodynamic limit ($M\rightarrow \infty$). 

\subsection{Weak confining potential derivations}
\label{weakdconferivations}

Consider edge excitations of form $\hat{L}^{\dagger}_{n}\ket{\Psi_{\rm L}} $, where $\hat{L}^{\dagger}_{n}  = \sum_{m=0}^{M} \alpha_{m}\hat{c}^{\dagger}_{m+n}\hat{c}_{m}$ and $\alpha_{m}=\bra{m+n}\hat{z}^{n}\ket{m} = (\sqrt{2}l_{B})^{n} \sqrt{\frac{(m+n)!}{m!}} $. Then their energy due to the confinement $\hat{V}_{M}$ is given by
\begin{equation}
E_{\Delta L = n} = \frac{\bra{\Psi_{\rm L}}\hat{L}_{n}\hat{V}_{M}\hat{L}^{\dagger}_{n}\ket{\Psi_{\rm L}}}{\bra{\Psi_{\rm L}}\hat{L}_{n}\hat{L}^{\dagger}_{n}\ket{\Psi_{\rm L}}}.
\end{equation} Proceeding similarly to the $n=1$ case, we use the fact that $\hat{V}_{M}\hat{L}^{\dagger}\ket{\Psi_{\rm L}}= [\hat{V}_{M},\hat{L}_{n}^{\dagger}]\ket{\Psi_{\rm L}}$ and $[\hat{V}_{M}, \hat{L}^{\dagger}_{n}] = V_{\text{conf}}\sum_{m=M-n+1}^{M}\alpha_{m}\hat{c}^{\dagger}_{m+n}\hat{c}_{m}$ to rewrite the energy of the edge state as 
\begin{equation}
\begin{aligned}
E_{\Delta L = n} = 
\frac{V_{\text{conf}}}{\bra{\Psi_{\rm L}}\hat{L}_{n}\hat{L}^{\dagger}_{n}\ket{\Psi_{\rm L}}} \sum_{m=M-n+1}^{M}|\alpha_{m}|^{2} \\ \times\left(\bra{\Psi_{\rm L}} \hat{c}^{\dagger}_{m}\hat{c}^{\dagger}_{m+n} \hat{c}_{m}\hat{c}_{m+n}\ket{\Psi_{\rm L}} +\braket{\hat{n}_{m}}\right), 
\end{aligned}
\end{equation}
where we have used properties of creation(annihilation) operators for further simplifications. 
We note that, in the last line, the first expectation value in the numerator vanishes as  $\hat{c}_{M+1}\ket{\Psi_{\rm L}}=0$. Plugging in the expression for $\alpha_{m}$ and rewriting the sum over $k=M-m$, we have that
\begin{equation}
\begin{aligned}
E_{\Delta L=n} = 
\frac{V_{\text{conf}}(2l_{B}^{2})^{n}}{\bra{\Psi_{\rm L}}\hat{L}_{n}\hat{L}^{\dagger}_{n}\ket{\Psi_{\rm L}}} \sum_{k=0}^{n-1}  \frac{(M-k+n)!}{(M-k)!}\braket{\hat{n}_{M-k}} \\
\approx \frac{V_{\text{conf}}(2l_{B}^{2})^{n} \braket{\hat{n}_{M}}}{\bra{\Psi_{\rm L}}\hat{L}_{n}\hat{L}^{\dagger}_{n}\ket{\Psi_{\rm L}}} M^{n} \sum_{k=0}^{n-1} \binom{k+q-1}{q-1}
\end{aligned}
\end{equation}
where we used the result\cite{mitra1993angular}
\begin{equation}
\braket{\hat{n}_{M-k}} \approx \binom{k+q-1}{q-1}\braket{\hat{n}_{M}},
\end{equation}
valid when $k\ll \sqrt{N}$ (i.e. when $n\ll \sqrt{N}$). The denominator also has an exact asymptotic expression\cite{dubail2012edge}
\begin{equation}
\bra{\Psi_{\rm L}}\hat{L}_{n}\hat{L}^{\dagger}_{n}\ket{\Psi_{\rm L}} \approx M^n (2l_{B}^{2})^{n} [n/q]
\end{equation}
and hence we obtain
\begin{equation}
\begin{aligned}
E_{\Delta L=n} \approx \frac{qV_{\text{conf}}}{ n}\braket{\hat{n}_{M}} \sum_{k=0}^{n-1}\binom{k+q-1}{q-1}\\
=\frac{qV_{\text{conf}}}{n}\braket{\hat{n}_{M}}\binom{n+q-1}{q}
.
\end{aligned}
\end{equation}
When $q=1$, the $n$ dependence vanishes, as is expected in the non-interacting case (flat dispersion and appearance of a gap). Moreover, for $n=1$, we recover the result presented in the paper. 

\subsection{Strong confining potential derivations}
\label{strongdconferivations}

\begin{figure}[htp]
\centering
\includegraphics[width=3.0in]{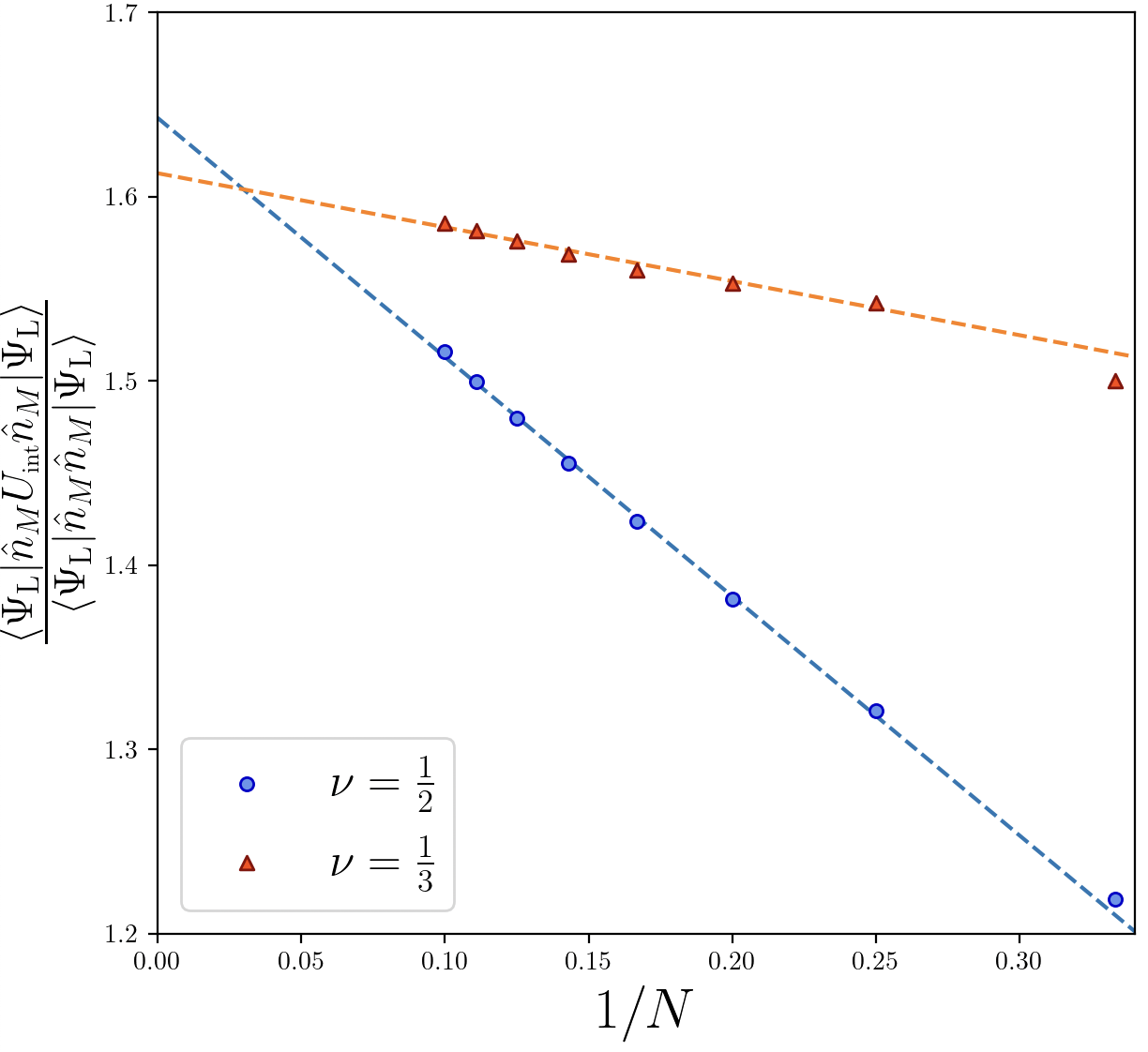}
\includegraphics[width=3.0in]{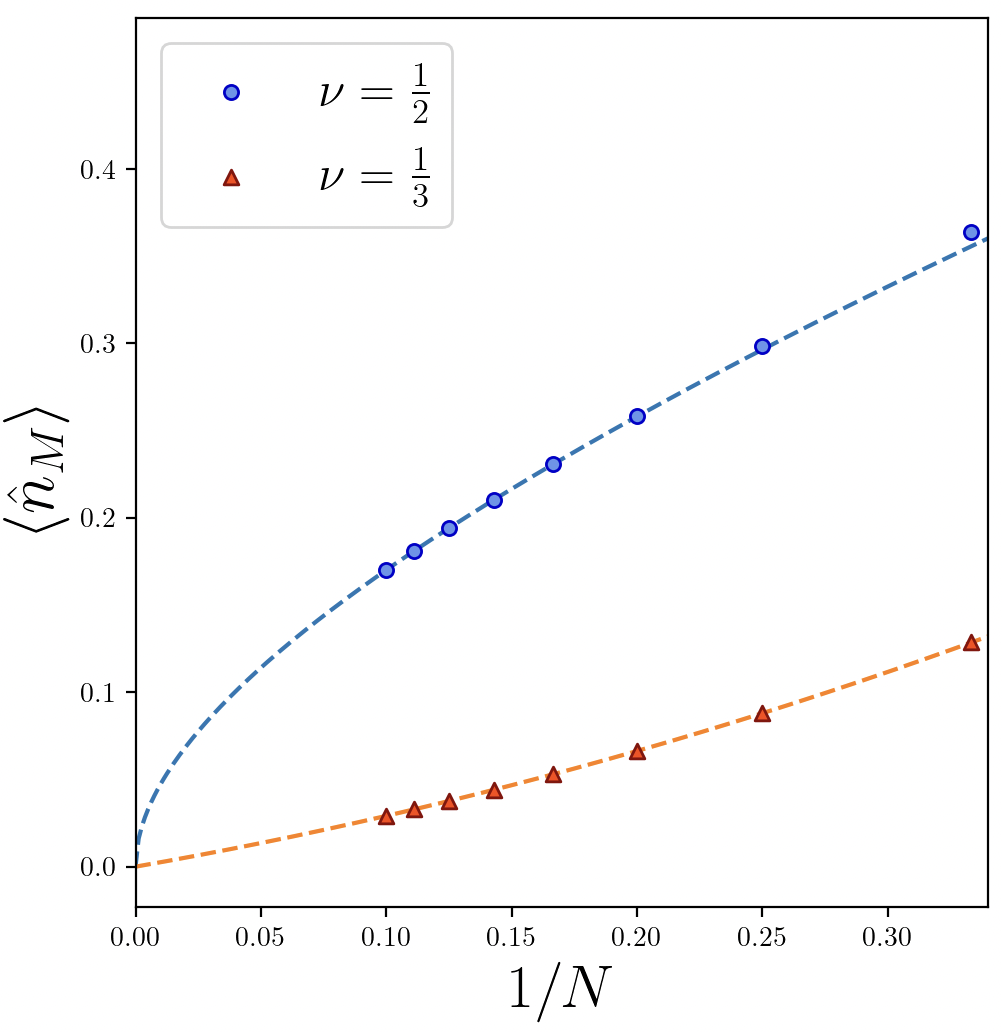}
\caption{(Upper) Expectation value of first factor in (\ref{EnergyTrialSCL}) as a function of system size. Fit given by $E_{0}(1-B N^{-1})$ for both $\nu=1/2$ and $\nu=1/3$. From fit, we obtain $E_{0} = 1.643 \pm 0.003$ for $\nu=1/2$ and $E_{0} = 1.613 \pm 0.003$. (Lower) Angular momentum occupation number of the Laughlin state $\braket{\hat{n}_{M}}$ as a function of system size. Fit given by $C N^{\frac{1-q}{2}} + D N^{1-q}$. From the fit, we obtain $b=0.443 \pm 0.003 $ for $\nu=1/2$ and $b=0.251 \pm 0.003 $ for $\nu=1/3$, giving an estimate of $f(q)$ in equation (\ref{SCLVariational}). }
\label{ProjectedEnergy+n_M}
\end{figure}

We present additional numerical data for the terms appearing in equation (\ref{EnergyTrialSCL}). The plot of the first factor as a function of system size is shown in Fig.~\ref{ProjectedEnergy+n_M} (upper).  Fitting $E_{0}(1-B N^{-1})$ shows that it tends to a finite value (of order ${\cal V}_{\rm int}$) as $N\to \infty$, as was argued in the main text. 
Furthermore, fitting also to $\braket{\hat{n}_{M}}$ in Fig.~\ref{ProjectedEnergy+n_M} (lower), we find a numerical estimate for $f(q)$ in equation (\ref{SCLVariational}). 
Combining these two results, we estimate the prefactor of the $N^{(1-q)/2}$ term in (\ref{SCLVariational}) to be $\approx 0.73\cal{V}_{\rm int}$ ($\nu=1/2$) and $\approx 0.40 \cal{V}_{\rm int}$ ($\nu=1/3$).

Finally, we note that the finite-size correction to $\bra{\Psi_{\rm L}} \hat{n}_{M}\hat{U}_{\text{int}}\hat{n}_{M}\ket{\Psi_{\rm L}}/{\bra{\Psi_{\rm L}} \hat{n}_{M}\hat{n}_{M}\ket{\Psi_{\rm L}}}$ and $\braket{\hat{n}_{M}}=C N^{\frac{1-q}{2}} + D N^{1-q}$ also introduce corrections to (\ref{EnergyTrialSCL}). We use the functional form $\frac{N^{\frac{1-q}{2}}}{1-A N^{\frac{1-q}{2}}}$ to account for these in the fits in Fig.~\ref{ThermoBosons+Fermions}.

\end{document}